\documentclass[conference]{IEEEtran}
\IEEEoverridecommandlockouts
\usepackage{cite}
\usepackage{amsmath,amssymb,amsfonts}
\usepackage{algorithmic}
\usepackage{graphicx}
\usepackage{textcomp}
\usepackage{xcolor}
\usepackage{fancyhdr}
\def\BibTeX{{\rm B\kern-.05em{\sc i\kern-.025em b}\kern-.08em
    T\kern-.1667em\lower.7ex\hbox{E}\kern-.125emX}}
\begin{document}

\title{Coarse-Fine View Attention Alignment-Based 
GAN for CT Reconstruction from Biplanar X-Rays}

\author{%
  \IEEEauthorblockN{%
    Zhi Qiao\IEEEauthorrefmark{1},
    Dongheng Chu\IEEEauthorrefmark{6},
    Hanqiang Ouyang\IEEEauthorrefmark{3}\IEEEauthorrefmark{4}\IEEEauthorrefmark{5},
    Huishu Yuan\IEEEauthorrefmark{3},
    Xiantong Zhen\IEEEauthorrefmark{1},
    Pei Dong\IEEEauthorrefmark{2},
    Zhen Qian\IEEEauthorrefmark{1}
  }%
  \IEEEauthorblockA{\IEEEauthorrefmark{1} Beijing United Imaging Research Institute of Intelligent Imaging, Beijing, China}%
  \IEEEauthorblockA{\IEEEauthorrefmark{2} United Imaging Intelligence
Beijing, China}%
  \IEEEauthorblockA{\IEEEauthorrefmark{3} Department of Radiology, Peking University Third Hospital, Beijing, China}%
  \IEEEauthorblockA{\IEEEauthorrefmark{4} Engineering Research Center of Bone and Joint Precision Medicine}%
  \IEEEauthorblockA{\IEEEauthorrefmark{5} Beijing Key Laboratory of Spinal Disease Research}%
  \IEEEauthorblockA{\IEEEauthorrefmark{6} Beijing University of Posts and Telecommunications, Beijing, China}%

}

\maketitle
\thispagestyle{fancy}
\fancyhead{}
\lhead{}
\lfoot{979-8-3503-3748-8/23/\$31.00 \copyright ©2023 IEEE}
\cfoot{}
\rfoot{}

\begin{abstract}
For surgical planning and intra-operation imaging, CT reconstruction using X-ray images can potentially be an important alternative when CT imaging is not available or not feasible. In this paper, we aim to use biplanar X-rays to reconstruct a 3D CT image, because biplanar X-rays convey richer information than single-view X-rays and are more commonly used by surgeons. Different from previous studies in which the two X-ray views were treated indifferently when fusing the cross-view data, we propose a novel attention-informed coarse-to-fine cross-view fusion method to combine the features extracted from the orthogonal biplanar views. This method consists of a view attention alignment sub-module and a fine-distillation sub-module that are designed to work together to highlight the unique or complementary information from each of the views. Experiments have demonstrated  the superiority of our proposed method over the SOTA methods. 
\end{abstract}

\section{Introduction}
X-ray imaging has been extensively used in orthopedic surgeries. X-rays provide excellent contrast for bones and have the advantages of lower radiation doses, lower expenses, and faster imaging speed over the other imaging technologies. However, a major limitation of X-rays is that they could not provide a 3D view of the internal structure. CT reconstruction using X-ray images can potentially be an important alternative when CT imaging is not available or not feasible. The 3D images generated through CT reconstruction from X-rays can provide valuable information for diagnosis and treatment planning. Additionally, the use of biplanar X-ray inputs, which are taken from two perpendicular angles, can further enhance the accuracy of the reconstruction, providing a more complete and detailed representation of the anatomy and structures of interest.

Some existing works have attempted 3D reconstruction from biplanar 2D X-ray images. In these biplanar X-ray-based CT reconstruction methods, a crucial step is the cross-view fusion, which is used to fuse the deep features derived from the two orthogonal biplanar views. The aforementioned methods use a matrix permutation process to keep the biplanar features geometrically consistent in orientation, followed by a direct addition or a concatenation operator for further fusion. The addition or concatenation process in the above cross-view fusion method assumes that each view provides an equivalent contribution to CT reconstruction. Nevertheless, the different views can demonstrate different morphological characteristics of the organs of interest. Intuitively, it is more beneficial to extract specific information from each view when performing cross-view fusion, because it helps to highlight unique or complementary information. Therefore, it is desirable to explore more efficient and effective fusion methods such that the relevant information from each view can be combined without unnecessary noise or redundancy.

In this paper, we propose a \textbf{C}oarse-Fine \textbf{V}iew \textbf{A}ttention \textbf{A}lignment-based GAN for CT reconstruction from biplanar X-rays (abbreviated as \textbf{CVAA-GAN}). Inspired by the advantages of GANs for cross-modality image transfer, especially in the medical domain, we adopt GAN as the main framework, which is composed of a Generator network and a Discriminator network. In the Generator, We propose the Coarse-Fine View Attention Alignment (CVAA) module to fuse the features extracted from the orthogonal biplanar views. More specifically, we first propose a view attention alignment sub-module to learn the fusion weights for cross-view fusion. Then, we introduce a fine-distillation sub-module to extract finer features from the coarse features. For the Discriminator, we opt to use a typical Patch-3D Discriminator.


\begin{figure*}[h]
\centering
\includegraphics[width=0.8\textwidth, height=0.35\textwidth]{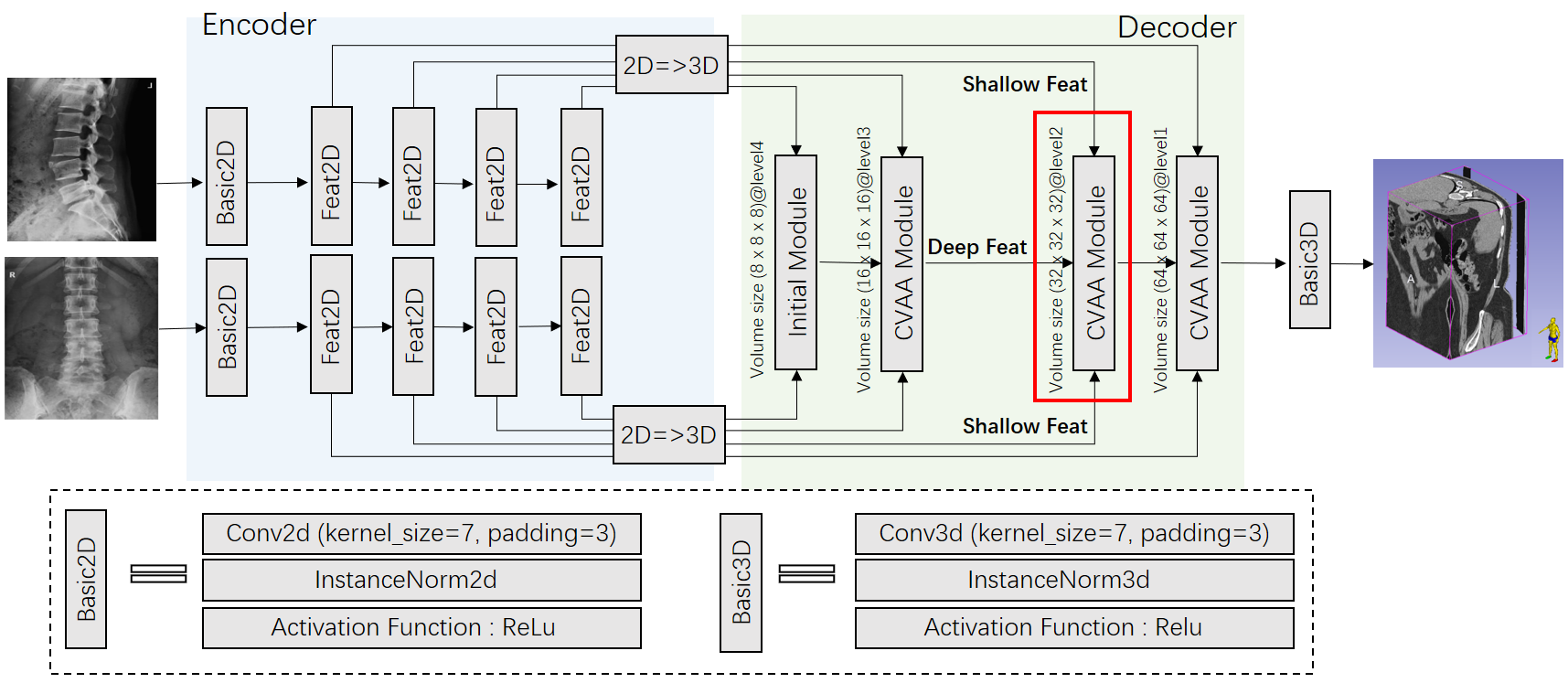}
\caption{Network architecture of the CVAA-GAN Generator}
\label{model:generator}
\vspace{-0.3cm}
\end{figure*}

\section{CVAA-GAN}
In this section, we introduce our proposed method. Similar to other 3D GAN architectures, our method involves a 3D Generator and a 3D Discriminator. These two models are alternatively trained.

\subsection{Generator}
The proposed 3D generator, as illustrated in Fig.\ref{model:generator}, consists of three individual components: two encoder networks to extract view features with the same architecture working in parallel for the two orthogonal views respectively, and a fusion decoder network for 3D image reconstruction.

\subsubsection{Encoder} 
The encoder network aims to extract features from the input 2D X-ray images and map them into 3D feature space.

\textbf{Feat2D}. To optimally utilize information from 2D X-ray images, we introduce the Feat2D modules in the generator’s encoding path where a typical densenet component is used. 

\textbf{2D$\Rightarrow$3D}. 
The 2D to 3D mapping procedure bridges the information from the 2D features to the 3D volumetric data. Motivated by \cite{4knee2xray}, we expand the 2D features to 3D by repeating the 2D feature maps along the X-ray's projection directions in 3D, which is based on an assumption commonly used by the back-projection-based CT reconstruction methods that the 3D features often repeat themselves along the projection direction. Then, we feed the initial 3D features into a basic 3D convolution block. In order to align the two 3D features in a consistent orientation, we transform them into a unified coordinate space.

\subsubsection{Decoder}
The decoder network is designed to fuse the biplanar view features, and project the deep features (lower resolution) learned by the encoder onto the voxel space (higher resolution) for final CT image reconstruction. We assume the biplanar X-ray images are captured within a negligible time interval, meaning no data shift caused by patient motions. In the deepest layer of the encoder, for initial low-resolution reconstruction, without any auxiliary information, we follow the existing view alignment method via the addition operator to combine these two view features to reconstruct the initial CT deep features (lowest resolution) which have the coarsest anatomical structure information, as shown in Fig. \ref{model:generator}. In the following, a basic 3D convolution block is used to enhance the feature correlation for the 3D structure, and then an up-convolution module is used to map the features into the higher-resolution feature space.

\textbf{CVAA Module}. In the following layers of the decoder, the inputs consist of two X-ray-related features from the encoders and the up-convolved features from the last decoder layer. The obtained X-ray-related features contain more high-frequency texture information, which can be seen as shallow features in the current resolution. The up-convolved features contain more low-frequency structural features which can be seen as deep features. The deep features can help identify which voxels are more useful and can guide a more reasonable fusion. In order to efficiently fuse the two views, we introduce a View Attention Alignment (VAA) sub-module to get the coarse features. Motivated by the coarse-fine scheme, we furthermore introduce a Fine Distillation (FD) sub-module to extract finer features from the coarse features. Besides the coarse and fine features, we also keep using the traditionally used view alignment features via the addition operator to enrich feature diversity, and the up-convolved features from the last decoder layer to take the advantage of the residual network. Finally, we ensemble all of the above features via a concatenation operator, followed by a basic 3D convolution layer and an up-convolution layer. After the up-convolution, the decoder outputs of the current layer are generated.

\begin{itemize}

\item View Attention Alignment. We first utilize the deep features to correlate two separate shallow features via a concatenating process and a basic 3D convolution block (Conv3d+Relu) to get the structure-aware view features. Two structure-aware view features are further concatenated and then fed into another basic Conv3D block (kernel size = 7, padding = 3, out\_channel = 2) to get the mixture features. Finally, a Softmax activation function is used to get attention weights. The weights are used to combine the two shallow features via operations of dot product and addition to derive the fusion features which are also considered the coarse features.  

\item Fine Distillation. After the coarse features are obtained, we further filter the noises and distill the coarse features. Firstly, we combine coarse features and deep features via concatenating to correlate coarse features with the structural information. Then, the combined features are fed into a basic Conv3D layer (kernel size = 3, padding = 1, out\_channel = 1), followed by a Sigmoid function to generate the distillation weights. The distillation weights are directly applied to the coarse features via dot product to derive the filtered features which are also considered the fine features.  

\end{itemize}
\begin{figure*}[h]
\centering
\includegraphics[width=0.87\textwidth, height=0.43\textwidth]{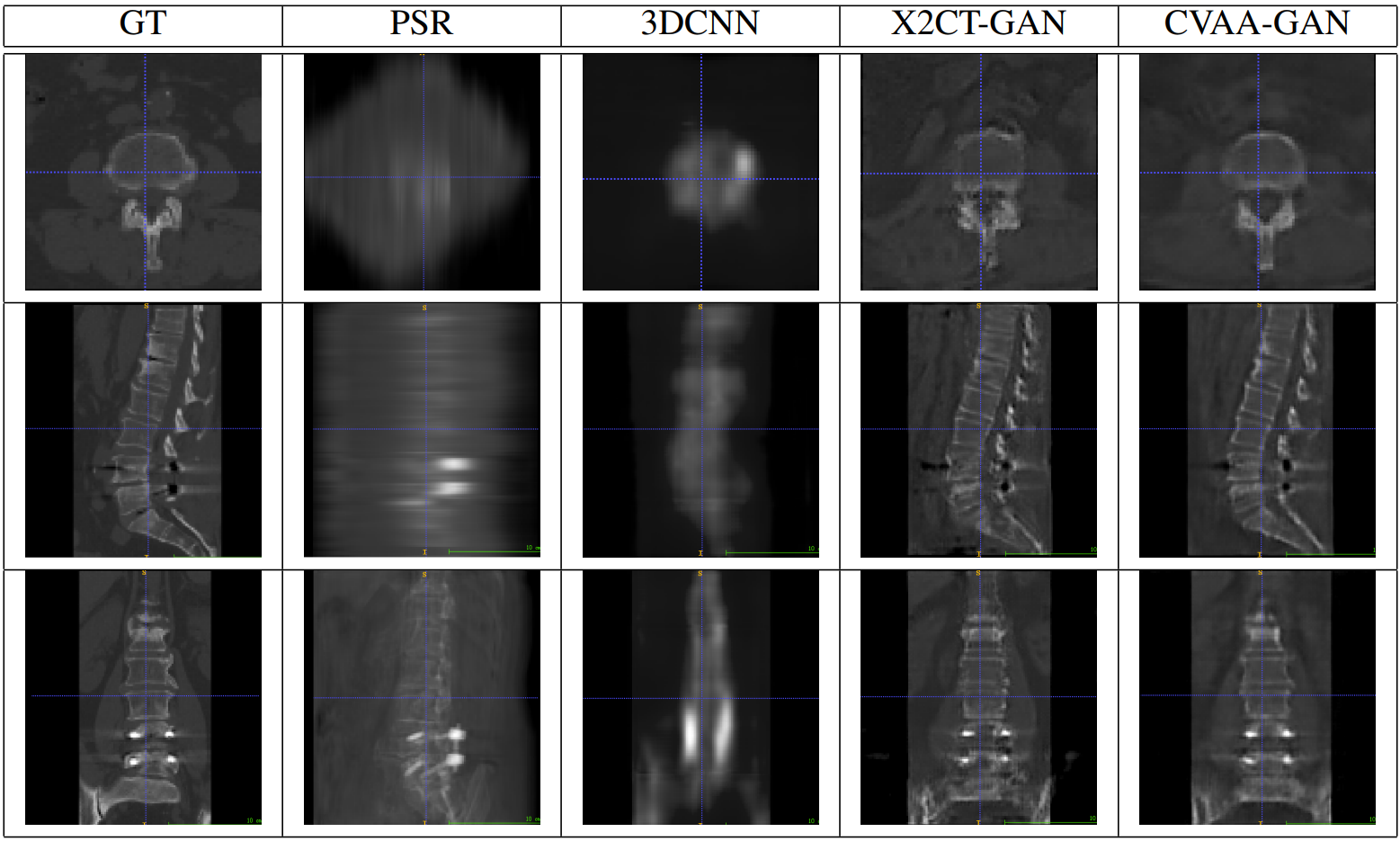}
\caption{Representative CT reconstruction results shown in the mid-axial (1st row), mid-sagittal (2nd row), and mid-coronal (3rd row) views. Our method is compared with baseline methods (PSR, 3DCNN, X2CT-GAN) and ground truth (GT).}
\label{exp:qualitative}
\end{figure*}

\subsection{Objective Functions}
In this section, we introduce the loss functions used in model learning.

\subsubsection{Adversarial Loss}
In image style translation, in contrast to Vanilla GAN and Wasserstein GAN \cite{8gan}, least squares generative adversarial networks (LSGANs) has been shown to generate higher quality images and perform more stable during the learning process \cite{12lsg}. In our framework, we adapt the loss function of LSGANs as the loss function for the discriminator. The objective functions can be defined as follows:
\begin{equation}
\begin{split}
&L(G) = E_X(||D(X, G(X)) - \mathbb{I}||^2) \\
&L(D) = E_X(||D(X, Y) - \mathbb{I}||^2) + E_X(||D(X, G(X)||^2)
\end{split}
\end{equation}
where $X = \{X_1, X_2\}$ are the input 2D Xray images, $Y$ is objective 3D CT scan, and $G(X)$ is the reconstructed 3D CT scan via the CVAA-GAN Generator. $E_X(\cdot)$ represents the expectation function, and here we use mean loss assumed to be independent and identically distributed. $D(X, Y)$ is used for conditional discrimination, where $X_1$ and $X_2$ are first extended from 2D to 3D and then followed by a basic 3D convolution layer (as the 2D$\Rightarrow$3D sub-module in the encoder pathway). Then, the 3D features combined with $Y$ via concatenation on the channel dimension are put into the discriminator. 

\begin{table*}
\scriptsize
\centering
\caption{Quantitative results of baseline methods and CVAA-GAN(The values in parentheses are the standard deviations.)}
\vspace{-2mm}
\begin{tabular}{|l|c|c|c|c|c|}
\hline 
Model & MAE & MSE& Cosine Similarity & PSNR & SSIM          \\ 
\hline 
PSR           & 0.03258(8.9e-5) & 0.003204(2e-6)& 0.9360(5e-4)& 25.1421(2.246)& 0.6025(0.002) \\
\hline 
3DCNN        & 0.02982(9.1e-5) & 0.003093(2e-6)& 0.9389(3e-4)& 25.4031(2.208)& 0.6328(0.001) \\
\hline 
X2CT-GAN      & 0.01975(4.6e-5) & 0.001765(8e-7)& 0.9664(1e-4)& 27.8361(2.947)& 0.7673(0.006) \\
\hline
CVAA-GAN    & \textbf{0.01635(3.1e-5)} & \textbf{0.001253(5e-7)}& \textbf{0.9757(7e-5)}& \textbf{29.7508(3.612)}& \textbf{0.8113(0.004)}
\\
\hline

\end{tabular}
\label{table:measurement}
\end{table*}

\subsubsection{Reconstruction Loss}
The conditional adversarial loss tries to make the prediction look similar to real data. However, it alone is not sufficient to ensure that the generated data is similar to the ground truth. Therefore, an additional constraint, such as the reconstruction loss, is required to enforce the reconstructed CT to be voxel-wise close to the ground truth. We define the reconstruction loss, 
$L = E_X(||Y-G(X)||^2)$
where, inspired by \cite{3dprojection}, we apply 2D projections to the predicted volume and compare the resulting projection views to the X-ray images from the corresponding ground truth. Orthogonal projections, instead of perspective projections, are carried out to simplify the process as this auxiliary loss focuses only on the general shape consistency, but not the X-ray veracity. We choose to use three orthogonal projection planes, i.e., the axial, the coronal, and the sagittal plane, respectively, following the convention in the medical imaging community. 
Some previous works have combined the reconstruction loss with the adversarial loss to improve the reconstruction performance. We also adopt this strategy. The adversarial loss plays an important role in encouraging local realism of the synthesized output. However, for surgical planning and intra-surgical observation, higher priority should be given to global and local shape consistencies.

\section{Experiments}
In this section, we first introduce a real-world lumbar vertebra dataset. We evaluate the proposed CVAA-GAN model with several commonly used metrics. To demonstrate the effectiveness of our method, we select several state-of-the-art methods as baselines. Fair comparisons and comprehensive analysis are given to demonstrate the improvement of our proposed method over the baselines. Finally, we show the CT reconstruction results from real-world biplanar X-rays using CVAA-GAN. 

\subsection{Datasets}
In the experiments, we use a privacy dataset to verify proposed method. A set of real-world lumbar vertebra postoperative CT data which is collected from a hospital. It contains 268 3D CT scans of 268 different patients. We randomly select 212 CT scans for training and the rest 56 CT scans are used for testing. We first resample the CT scans to the same isotropic voxel resolution ($2\times 2 \times 2 ~ mm^3$), followed by center-cropping with a fixed crop size ($128 \times 128 \times 128$). We then clip data with CT value span [-1000, 4096], and use min-max normalization to scale data into 0-1 range. Ideally, paired CT and biplanar X-ray data are needed to train and test our proposed model. However, such a paired dataset is difficult to collect and currently not available. Therefore, we opt to use the digitally reconstructed radiographs (DRR) technology \cite{13ddr} to synthesize the corresponding X-rays from volumetric CT images and generate the paired CT and X-ray images. 

\subsection{Settings}
The generator and discriminator are trained alternatively following the standard GAN process. Meanwhile, we use instance normalization to regularize intermediate feature maps of our generator \cite{10norm}. We use the Adam solver \cite{9adam} to train our networks. The learning rate of Adam is 2e-4, momentum parameters $\beta_1 = 0.5$ and $\beta_2 = 0.99$. We train our model for a total of 200 epochs. All of methods are conducted in NVIDIA A40. Constrained by GPU memory limit, the batch size is set to 4 in all our experiments. Three state-of-the-art methods are selected as baselines for CT reconstruction task, PSR \cite{1Singleimage}, 3DCNN \cite{4knee2xray}, X2CT-GAN \cite{5X2CT}. For the baselines, we reproduce PSR, 3DCNN, and use the open-source codes of X2CT-GAN. The parameters of models are set as in publications.  

\subsection{Qualitative Results}
We first qualitatively evaluate CT reconstruction results as
shown in Fig.\ref{exp:qualitative}. For each method, the mid-axial, mid-sagittal, and mid-coronal slices of the reconstruction CT are shown in the first, second, and third rows, respectively. PSR just using a single-view X-ray generates blurrier volumes compared to the other biplanar X-ray-based models. Compared with 3DCNN, X2CT-GAN, and ours generate sharper boundaries of the organs and implants, suggesting the cross-view fusion in the decoder is effective in preserving more fine-grained information from the input X-ray data. Last, our proposed CVAA-GAN visually performs the best, generating the closest morphological characteristics of the vertebrae and the implants to the ground truth, and the sharpest boundaries, suggesting our method achieves less reconstruction distortion and less detailed texture loss.

\subsection{Quantitative Results}
Quantitative results are summarized in Table \ref{table:measurement}.
We can find that 1) PSR with single X-ray input has the worst performance compared to other biplanar models. 2) X2CT-GAN and our model have better performance than 3DCNN.     The cross-view fusion in the decoder pathway is essential in recovering the underlying 3D anatomy. 3) Compared with the best baseline, CVAA-GAN shows apparent performance improvement, suggesting the view attention alignment sub-module can leverage different strengths of different view features in the cross-view fusion step and the fine distillation sub-module can direct the model to focus on more valuable information that is complementary to each of the views without unnecessary noise or redundancy.

\section{Conclusions}
In this paper, we propose a novel view attention alignment to fuse biplanar X-ray features which can leverage different characteristics from different views. Furthermore, we introduce a coarse-to-fine scheme to extract finer features from coarse features to further improve algorithm performance. Experimental results demonstrated the superior performance of our approach compared with the SOTA baselines.

\section*{Acknowledgment}
The work is supported by the following projects, National Natural Science Foundation of China (82171927), Peking University Third Hospital Clinical Key Project (BYSY2018003), National Natural Science Foundation of China (82102638) and Peking University Third Hospital Clinical Key Project- (BYSYZD2021040).

\bibliographystyle{IEEEtran} 
\bibliography{main}

\end{document}